\documentclass[aps,pre,twocolumn,floatfix]{revtex4}
\usepackage{graphicx}
\usepackage{epsfig}

\newcommand{\be}{\begin{equation}}
\newcommand{\ee}{\end{equation}}

\begin{document}

\title[Azimuthal asymmetries of the LSC in turbulent RB convection]{Azimuthal asymmetries of the large-scale circulation \\ in turbulent Rayleigh-B{\'e}nard convection}
\author{Eric Brown}
\altaffiliation[Present address:  ]{The James Franck Institute, University of Chicago, Chicago, IL 60637}
\author{Guenter Ahlers}
\affiliation{Department of Physics and iQCD, University of California, Santa Barbara, CA 93106}
\date{\today}

\begin{abstract}

Previously we published a dynamical model (E. Brown and G. Ahlers, Phys. Fluids, {\bf 20}, 075101 (2008)) for the large-scale-circulation (LSC) dynamics of Rayleigh-B{\'e}nard convection in cylindrical containers. The model consists of a pair of stochastic ordinary differential equations, motivated by the Navier-Stokes equations, one each for the strength $\delta$ and the orientation $\theta_0$ of the LSC. Here we extend it to cases where the rotational invariance of the system is broken by one of several physically relevant perturbations.  As an example of this symmetry breaking we present experimental measurements of the LSC dynamics for a container tilted relative to gravity. In that case the model predicts that the buoyancy of the thermal boundary layers encourages fluid to travel along the steepest slope, that it locks the LSC in this direction, and that it strengthens the flow,  as seen in experiments.   The increase in LSC strength is shown to be responsible for the observed suppression of cessations and azimuthal fluctuations.   We predict and observe that for large enough tilt angles, the restoring force that aligns the flow with the slope is strong enough to cause oscillations of the LSC around this orientation.  This planar oscillation mode is different from coherent torsional oscillations that have been observed previously.   The model was applied also to containers with elliptical cross-sections and predicts that the pressure due to the side walls forces the flow into a preferred orientation in the direction of the longest diameter. When the ellipticity is large enough, then oscillations around this orientation are predicted.  The model shows that various azimuthal asymmetries will lock the LSC orientation. However, only those that act on the $\delta$-equation (such as tilting relative to gravity) will  enhance the LSC strength and suppress cessations and other azimuthal dynamics. Those that affect only the $\theta_0$ equation, such as an interaction with Earth's Coriolis force, will align the flow but will not influence its strength and the frequency of cessations.

\end{abstract}
 

 \maketitle
 
\section{Introduction}

The problem of Rayleigh-B{\'e}nard convection (RBC) consists of a fluid sample heated from below and cooled from above (for reviews, see Refs. \cite{Si94, Ka01, AGL02}).  In turbulent convection, a hot (cold) thermal boundary layer at the bottom (top) becomes unstable due to buoyancy and emits hot (cold) volumes of fluid known as ``plumes" which detach from the boundary layer and transport heat vertically.  In cylindrical containers with aspect ratio $\Gamma \equiv D/L \approx 1$ ($L$ is the height and $D$ is the diameter of the sample) these plumes contribute to the driving and are carried by a large-scale circulation (LSC), also know as the mean wind,  which forms a loop with up-flow and down-flow on opposite sides of the sample \cite{HCL87, SWL89, CGHKLTWZZ89, CCL96, TSGS96, CCS97, QT00, QT01a, QT01b, NSSD01, QT02, QSTX04, FA04, SXT05,TMMS05}.    The LSC breaks the rotational invariance of a cylindrical sample and its near-vertical circulation plane must somehow choose an azimuthal orientation.  This orientation has been found to undergo spontaneous diffusive meandering \cite{SXX05, XZX06, BA06a, BA06b}.   The LSC also undergoes re-orientations both by azimuthal {\em rotations} \cite{CCS97, BA06a}, and by {\em cessations} in which it essentially slows to a stop and then restarts in a near-random new orientation \cite{BNA05,BA06a}.   The dynamics of the LSC also are known to include torsional oscillations in which the orientation of the upper half oscillates out of phase with the lower half \cite{FA04, ZSX07, RPTDGFL06, FBA08}.  

For cylindrical containers much of the LSC dynamics are described very well by a model presented in Ref.~\cite{BA07a} and studied in detail in Ref.~\cite{BA08a} (an exception is the torsional oscillation; although it is known that this mode has the properties of a damped oscillator  driven by the turbulent background fluctuations\cite{FBA08}, its origin remains obscure at this time).  This model was physically motivated by the Navier-Stokes (NS) equations and was reduced to a pair of stochastic ordinary differential equations.  The LSC strength was represented by a temperature amplitude $\delta$.  The change in $\delta$ is determined by the balance between buoyancy and drag forces, and the system is driven by stochastic fluctuations $f_{\delta}(t)$ which phenomenologically represent the action of the small-scale turbulent background fluctuations on the LSC:

\be
\dot\delta = \frac{\delta}{\tau_{\delta}} - \frac{\delta^{3/2}}{ \tau_{\delta}\sqrt{\delta_0}} + f_{\delta}(t) \ .
\label{eq:lang_delta}
\ee

\noindent Here $\delta_0$ is the fixed-point amplitude of the deterministic equation and is nearly equal to the mean value of $\delta$ expected to be observed in experiment. The azimuthal acceleration of the LSC orientation $\theta_0$ is damped by the rotational inertia of the LSC and driven by stochastic fluctuations, again representing the turbulent fluctuations:

\be
\ddot\theta_0 = - \frac{\dot\theta_0\delta}{\tau_{\dot\theta}\delta_0} + f_{\dot\theta}(t) \ .
\label{eq:lang_theta}
\ee

\noindent The stochastic terms were assumed to be Gaussian distributed white noise, leading to  experimentally measured diffusivities $D_{\delta}$ and $D_{\dot\theta}$.  The timescales were predicted by the model to be 
$\tau_{\delta} = L^2/(18\nu R_e^{1/2})$ and $\tau_{\dot\theta} = L^2/(2\nu R_e)$ where $\nu$ is the kinematic viscosity of the fluid and $R_e$ is the Reynolds number.  Measurements of the LSC model parameters from the same experimental apparatus as the present measurements were reported in Ref.~\cite{BA08a} (for the diffusivities and timescales) and Ref.~\cite{BFA07} (for the Reynolds number).

The goal of the present work was to study the LSC dynamics in systems with slight azimuthal asymmetries to determine if and how results found in cylindrical containers can be applied to such asymmetric systems.   This is an important extension since most natural convection systems do not have the symmetry of ideal laboratory experiments.  One such asymmetry is Earth's Coriolis force, which was found to cause a net rotation of the LSC orientation on average once every 3 days, and to align the LSC in a preferred orientation close to West \cite{BA06b}.  One natural convection system with azimuthal asymmetry is Earth's outer core which  drives Earth's magnetic field with orientation changes\cite{GCHR99} which some believe could be caused by cessations.  Another is the solar convection zone which has both torsional oscillations and periodic reversals \cite{HL80b}.  Both of these examples are spherical geometries which become asymmetric due to the action of external fields.  
Reversals are also known to occur in the wind direction in the atmosphere \cite{DDSC00}.

In Sect.~\ref{sect:tilt}  we extend the model of the LSC dynamics to the case where the sample is tilted slightly relative to gravity.  It is known from experiments that tilting the sample breaks the rotational symmetry significantly and aligns the flow along the direction of tilt with the flow going up along the hot bottom plate and down along the cold top plate \cite{CCL96,ABN06}.  It has been shown that the Reynolds number \cite{CRCC04, ABN06} and average amplitude of the LSC \cite{ABN06} increase with tilt due to the component of  buoyancy of fluid in the thermal boundary layers parallel to the plates, and that the average frequencies of occurrence of both cessations and rotations are reduced \cite{BA06a}.  The orientation-locking effect is due to a component of the same buoyancy that aligns the flow with the steepest slope of the plates \cite{BA06a}.  The models of Refs.~\cite{CRCC04, ABN06, BA06a} used a steady-state force balance between buoyancy and drag in the boundary layers.  Here these forces are generalized to the dynamic case so they can be incorporated in the dynamical model of Ref.~\cite{BA07a}.  In Sect.~\ref{sect:tilt_experiment} the model predictions for a tilted sample are compared with experimental results. The good agreement provides further support for the model of Ref.~\cite{BA07a}.  We also report a prediction of a new in-phase azimuthal oscillation due to the azimuthal asymmetry, and the experimental observations of this mode, in a tilted container.  Section \ref{sect:noncircular} contains an extension of the model to non-circular containers and comparisons with the deterministic model of Ref.~\cite{RPTDGFL06}.  Section \ref{sect:coriolis} contains a discussion of the experimental results of Ref.~\cite{BA06b} on the effects of Earth's Coriolis force and of asymmetric heating, and how those results relate to the stochastic model.

Finally we note that several other models of the LSC have been presented in the literature. We mentioned four of them in a previous paper \cite{NS03,FGL05,Be05,RPTDGFL06} and discussed how  they differ from our model; we shall not repeat this discussion at this point. However, very recently yet another model was developed \cite{RJH08a,RJH08b} for a system with a square cross section of aspect ratio four with periodic lateral boundary conditions. It considers the coupling between the temperature and velocity field, but has no azimuthal degree of freedom. It is a deterministic model with relaxational dynamics, and thus can not reproduce any of the statistically stationary dynamics (which include fluctuation amplitudes and cessations) of our model and of the  physical system.

\section{The experiment}

Measurements were taken in the cylindrical sample described in detail as the medium sample of Ref.~\cite{BNFA05}.  The top and bottom plates were made of copper and the side wall was made of plexiglas. The sample had height $L = 24.76$ cm and diameter $D =24.81$ cm, yielding an aspect ratio $\Gamma = 1.00$.   The container was surrounded by a thermostatic shield kept at the fluid temperature at the mid-height of the sample to minimize heat flux through the side of the sample.  The working fluid was water at a mean temperature of $40.00^\circ$C.  The Prandtl number $\sigma$, defined by

\be
\sigma \equiv \frac{\nu}{\kappa}\ , 
\label{eq:prandtl}
\ee

\noindent ($\kappa$ is the thermal diffusivity of the fluid) was equal to 4.38  ($\nu = 6.7\times 10^{-7}$ m$^2/$s) for our experiments.  The temperature difference $\Delta T$ between the top and bottom plates was controlled with a resistive heater imbedded in the bottom plate and cooled with water pumped through a channel in the top plate.   The Rayleigh number $R$ is a control parameter  defined by

\be
R \equiv \frac{\alpha g \Delta T L^3}{\kappa \nu}
\label{eq:rayleigh}
\ee

\noindent  where $\alpha$ is the isobaric thermal expansion coefficient and $g$ is the acceleration of gravity.   We ran experiments with different $\Delta T$ to obtain a Rayleigh number range of $2\times 10^8 \stackrel{<}{_\sim} R \stackrel{<}{_\sim} \times10^{10}$.   The other control parameter was the tilt angle $\beta$ of the sample relative to gravity, which was changed by turning a leveling screw to vary the extent of the azimuthal asymmetry of the system.

The LSC was observed using thermistors imbedded in the side wall from the outside so they were near to but not penetrating into the fluid region \cite{ABN06,FBA08}.   The interior of the container had no measuring devices or other objects that could break the azimuthal symmetry of the system.  Depending on $R$, the dominating azimuthal asymmetry for the level-sample experiments was either due to Earth's Coriolis force or to a small asymmetry in the top-plate cooling system \cite{BA06b}.  Eight  thermsitors were placed, equally spaced azimuthally, at each of the three heights $-L/4$, $0$, and $L/4$ (the origin of the vertical axis is taken to be at the horizontal midplane of the sample).  The sampling period was about 2.5 seconds.  Since the LSC carried warm fluid from the bottom plate up one side and cold fluid down the other side, the temperture profile indicated the orientation and strength of the LSC.  The thermistor readings were  fit, separately at each height and time step, by the function 

\be
T = T_0 + \delta\cos\left[\theta - \theta_0\right]\ .
\label{eq:temp_profile}
\ee

\noindent Of the  three fit parameters -- the temperature amplitude $\delta$, the mean temperature $T_0$, and the azimuthal orientation $\theta_0$ -- $\delta$ and $\theta_0$  are the same two variables that  were used in the stochastic model to describe the LSC.  As defined here, the orientation $\theta_0$ is on the side of the sample where the LSC is warm and up-flowing.  Results from this method of determining the orientation and strength of the LSC were reported  in several previous publications \cite{BNA05,ABN06,BA06a,BA06b, XX07, XX07b, BA07b, BFA07, FBA08, BA08a}.  Measurements of the LSC orientation and amplitude reported in this paper were taken at the mid-height unless otherwise noted.  The measurements at all 3 heights are necessary to measure and distinguish between the torsional oscillation mode \cite{FA04} and the in-phase oscillations that we found due to tilting the sample.

\section{The model for tilted containers}
\label{sect:tilt}

\subsection{The model equations}
\label{sec:model_equations}

We now consider how the LSC dynamics change when the sample is tilted slightly relative to gravity with the high side at the orientation $\theta_{\beta}$.   The additional forces that are not present in the level sample are formulated as volume averages so that they can be added as perturbative terms to the model represented by Eqs.~\ref{eq:lang_delta} and \ref{eq:lang_theta}.  By tilting the sample, the buoyancy of the thermal boundary layers is no longer aligned with the cylinder axis, but a component $\sin\beta$ of that force is along the direction of tilt $\theta_{\beta}$.  Fluid in the bottom boundary flows up-slope, while fluid in the top boundary flows down-slope, so their contributions add constructively to enhance and align the LSC in the direction of $\theta_{\beta}$. The additional acceleration in the direction parallel to the plates is given by the Navier-Stokes equation to be $\dot u' = g\alpha (T-T_0) \sin\beta$.   Approximately half of the temperature drop in the sample occurs in each boundary layer, so the average temperature of each of the thermal boundary layers relative to the mean is $\Delta T/4$.  The total volume fraction of the boundary layers is $2l/L$, where $l = L/(2\mathcal{N})$  is an adequate approximation for the thermal boundary-layer width \cite{GL02} ($\mathcal{N}$ is the Nusselt number).  Multiplying these terms gives the volume-averaged acceleration 

\be
\langle \dot u'\rangle_V = \frac{g\alpha \Delta T }{4\mathcal{N}} \sin\beta \ .
\label{eq:tilt_forcing}
\ee

\noindent A fraction $\cos(\theta_0-\theta_{\beta})$ of this acceleration is in the direction of the LSC and adds to the LSC strength when it is aligned with $\theta_{\beta}$, and a fraction $\sin(\theta_0-\theta_{\beta})$ pushes in the azimuthal direction towards $\theta_{\beta}$.  The model describes the LSC strength in terms of the measured parameter $\delta$, so we use the assumption that $\delta \propto U$ \cite{BA07a} where $U$ is the maximum velocity in the coordinate of the LSC (which occurs at the edge of the bulk region just outside of the boundary layers) to get

\be
\dot\delta_{tilt} = \frac{\delta_0\langle \dot u'\rangle_V}{\langle U\rangle}\cos(\theta_0-\theta_{\beta}) \ .
\label{eq:tilt_delta}
\ee

\noindent The azimuthal rotation rate $\dot\theta_0$ is related to the azimuthal velocity $u_{\theta}$  by $\dot\theta_0 = u_{\theta}/r$ where $r$ is the radius in cylindrical coordinates.  The volume average for $\ddot\theta_0$ becomes 

\be
\ddot\theta_{tilt} = \left<\frac{\dot u_{\theta}}{r}\right>_V = \frac{3\langle \dot u'\rangle_V}{L}\sin(\theta_0-\theta_{\beta}) \ .
\label{eq:tilt_theta}
\ee

\noindent This calculation differs somewhat from those of Refs.~ \cite{CRCC04, ABN06} because in those cases the force balance between buoyancy and drag was made only in the viscous boundary layer.  Following the spirit of Ref.~\cite{BA07a}, here volume averages were taken, which allowed a consideration of buoyancy in the thermal boundary layer separate from the drag in the viscous boundary layer.  These differences do not affect the qualitative dynamics of the model, but they do change numerical coefficients by factors of order 1 and influence slightly the dependence of the predictions on the Rayleigh number. The amplitude equation Eq.~\ref{eq:lang_delta} with the addition of the tilt term of Eq.~\ref{eq:tilt_delta} becomes

\be
\dot\delta = \frac{\delta}{\tau_{\delta}} - \frac{\delta^{3/2}}{\tau_{\delta}\sqrt{\delta_0}} + f_{\delta}(t) + \delta_{\beta}\sin\beta\cos(\theta_0-\theta_{\beta})\ .  
\label{eq:lang_tilt_delta}
\ee

\noindent Using the definitions of $R$ (Eq.~\ref{eq:rayleigh}), $\sigma$ (Eq.~\ref{eq:prandtl}), and the Reynolds number  $R_e=\langle U\rangle L/\nu$ we find

\be
\delta_{\beta} = \frac{R\nu \delta_0}{4\mathcal{N}R_e \sigma L^2}\ .
\label{eq:dot_delta_beta}
\ee

\noindent The azimuthal equation Eq.~\ref{eq:lang_theta} with the addional term of Eq.~\ref{eq:tilt_theta} becomes

\be
\ddot\theta_0 = -\frac{\delta\dot\theta_0}{\delta_0\tau_{\dot\theta}} + f_{\dot\theta}(t) - \omega_{\beta}^2 \sin\beta\sin(\theta_0-\theta_{\beta}) 
\label{eq:lang_tilt_theta}
\ee

\noindent where

\be
\omega_{\beta}^2 = \frac{3\nu^2 R}{4\mathcal{N}\sigma L^4} \ .
\label{eq:omega_beta}
\ee

\noindent Equations \ref{eq:lang_tilt_delta} and \ref{eq:lang_tilt_theta} compose the model of the LSC dynamics in tilted containers.  Most of the data presented in this paper are taken at $R=2.8\times 10^{8}$.  For this $R$,  it was found that $\delta_0 =0.020$ K, $\tau_{\delta} = 133$ s, $D_{\delta} = 8.7\times10^{-8}$ K$^2$/s, $\tau_{\dot\theta} =  14 $ s, and $D_{\dot\theta} = 1.5\times10^{-6}$ rad$^2$/s$^3$ using the methods of Ref.~\cite{BA08a}.  Making additional use of measurements of $\mathcal{N}$ \cite{NBFA05} and $R_e$ \cite{BFA07} from the same apparatus, the predictions for the tilt parameters at this $R$ are $\delta_{\beta} = 1.2\times10^{-4}$ K/s and $\omega_{\beta}^2 = 1.3\times10^{-4} $ s$^{-2}$. 
The measured values of $\delta_{\beta}$ and $\omega_{\beta}^2$ are reported in Sect.~\ref{sect:tilt_experiment}.

\subsection{The potential}

The stochastic amplitude equation can be understood as describing the diffusion of $\delta$ in a potential $V$ defined by $V \equiv -\int \dot\delta_{det} d\delta$ where $\dot\delta_{det}$ is the deterministic part of Eq.~\ref{eq:lang_tilt_delta}. This gives

\be
V  = - \frac{\delta^2}{2\tau_{\delta}} + \frac{2\delta^{5/2}}{5\tau_{\delta}\sqrt{\delta_0}} - \delta_{\beta}\delta \sin\beta\cos(\theta_0-\theta_{\beta})\ . 
\label{eq:potential}
\ee

\begin{figure}                                                
\centerline{\includegraphics[width=3.25in]{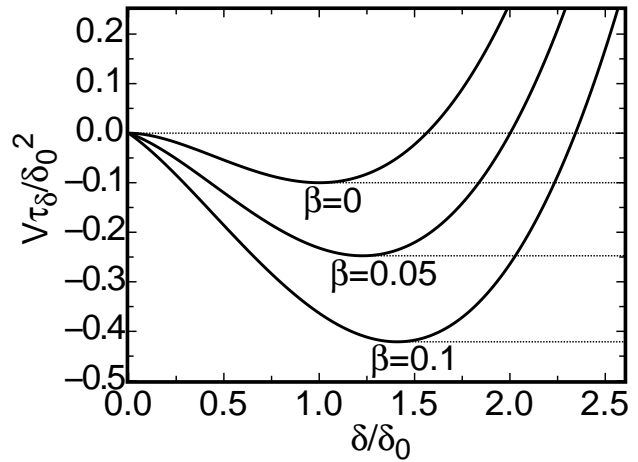}}
\caption{The potential $V = -\int \dot\delta_d d\delta$ from Eq.~\ref{eq:potential} for three tilt angles at $R=1.1\times10^{10}$ and $\theta_0 = \theta_{\beta}$. The potential is non-dimensionalized by $\tau_{\delta}/\delta_0^2$ so the $\beta=0$ curve is the same for all $R$, but the value of $\dot\delta\tau_{\delta}/\delta_0$ and hence the magnitude of the shift with $\beta$ varies with $R$.   The displacements relative to zero of the horizontal dotted lines indicate the heights of the potential barriers $\Delta V$ for cessations for each tilt angle. }
\label{fig:potential}                                       
\end{figure}

\noindent This potential has the form of a well with a stable fixed point where $\dot\delta(\delta)=0$. For $\beta=0$ this point occurs at $\delta=\delta_0$ .  The potential goes to infinity for large $\delta$, but has the finite value $V = 0$ at $\delta=0$.  When fitting the temperature profile Eq.~\ref{eq:temp_profile} to the data,  $\delta$ is forced to be positive definite so that $\delta$ and $\theta_0$ can be determined uniquely. Thus, if $\delta$ approaches zero, then this implies that the LSC is about to reverse direction.
These reversals are handled with a reflective boundary condition at $\delta=0$, and upon reflection $\pm\pi$ is added to $\theta_0$.  This is equivalent to a double-well potential in $\delta$ that is symmetric around $\delta=0$.  The potential is shown in Fig.~\ref{fig:potential} for three tilt angles. The minimum of the potential depends quite strongly on $\beta$.  It is immediately clear that cessations will be strongly suppressed by a tilt with the deeper minimum because they depend on diffusion, driven by the background fluctuations, from the neighborhood of the potential minimum to $V=0$.

\subsection{The linearized model}

For ease of calculations, the model can be linearized by expanding around the stable fixed point at the minimum of the potential well with the substitution $\epsilon\equiv\delta - \delta_0$ as was done in Ref. \cite{BA08a}.  Since $\delta$ fluctuates around the potential minimum, this is a good approximation for calculations of most average values, but it will not work for studying cessations which occur when $\delta$ becomes small.  It is known from experiments that the orientation $\theta_0$ becomes restricted to a narrow range of $\theta$ near $\theta_\beta$ even for very small tilt angles \cite{BA06a}.  For the smallest measured tilt angle $\beta=0.017$ rad, the orientation was confined to a range of width $0.38$ rad around $\theta_{\beta}$ so that $ \langle\cos(\theta_0-\theta_{\beta})\rangle \approx 0.93$.  Thus, setting the cosine term equal to 1 underestimates the tilt term by only 7\%, and this error diminishes rapidly for larger tilt angles where the orientation is confined closer to $\theta_{\beta}$.  Keeping only the lowest-order term in $\epsilon$, the linearized version of Eq.~\ref{eq:lang_tilt_delta} is then

\be
\dot\epsilon = -\frac{\epsilon}{2\tau_{\delta}} + f_{\delta}(t) + \delta_{\beta}\sin\beta  \ .
\label{eq:lang_delta_lin}
\ee

\noindent This equation is valid for tilt angles large enough to lock the orientation but small enough that the fixed point amplitude $\bar\delta$ remains in the parabolic region of the $\beta=0$ potential.  The fixed-point amplitude in this approximation is given by

\be
\bar\epsilon = 2\tau_{\delta}\delta_{\beta}\sin\beta \  .  
\label{eq:bar_epsilon}
\ee

\noindent The linearized potential is defined by $V_{\epsilon}(\epsilon) \equiv V(\delta_0) -\int \dot\epsilon_{det} d\epsilon$ where $\dot\epsilon_{det}$ is the deterministic part of Eq.~\ref{eq:lang_delta_lin}. This gives

\be
V_{\epsilon}(\epsilon) = \frac{\epsilon^2}{4\tau_{\delta}} -(\delta_0 +\epsilon) \delta_{\beta}\sin\beta -\frac{\delta_0^2}{10\tau_{\delta}}\ .
\label{eq:potential_lin}
\ee

The azimuthal equation Eq.~\ref{eq:lang_tilt_theta} can be linearized for small $\theta_* \equiv \theta_0-\theta_{\beta}$. Setting $\delta$ equal to the constant fixed point value $\bar\delta$ one gets

\be
\ddot\theta_* = -\frac{\bar\delta\dot\theta_*}{\delta_0\tau_{\dot\theta}} - \omega_{\beta}^2 \sin\beta ~\theta_*+f_{\dot\theta}(t) \ .
\label{eq:lang_theta_lin}
\ee

\noindent The fluctuations in $\delta$ are relatively small (see Fig.~\ref{fig:sigma_delta}) so ignoring them will not affect the calculations of most average values, but again the linearized equation will not be valid for cessations when $\delta$ becomes small.

\section{Results for tilted containers}
\label{sect:tilt_experiment}

\subsection{The amplitude $\delta$}

\begin{figure}                                                
\centerline{\includegraphics[width=3.25in]{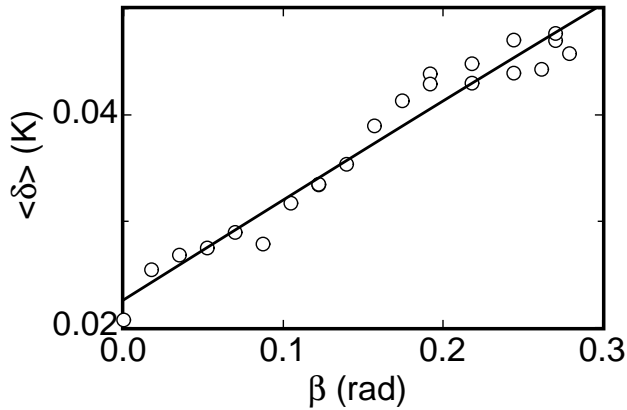}}
\caption{The average amplitude $\langle\delta\rangle$, which is used as a measure of the fixed-point amplitude $\bar\delta$, as a function of tilt angle $\beta$ at $R=2.8\times10^8$. The solid line is a linear fit to the data.}
\label{fig:delta_tilt}                                       
\end{figure}

The time-averaged measured amplitude $\langle\delta\rangle$ is shown in Fig.~\ref{fig:delta_tilt} for various tilt angles at $R=2.8\times10^8$.  Since $\delta$ fluctuates in a near-parabolic well, $\langle \delta\rangle$ is a good approximation to the fixed-point amplitude $\bar\delta$. Consistent with Eq.~\ref{eq:bar_epsilon}, $\langle \delta \rangle$ varies linearly with $\beta$ at these small values of $\beta$ where $\sin \beta \simeq \beta$.  A fit of $\langle\delta\rangle = \delta_0 + A\sin\beta$ gave $A=0.094$ K.  This should be compared with the prediction of $A = 2\tau_{\delta}\delta_{\beta} \approx 0.031$ K from Eq.~\ref{eq:bar_epsilon}, which is about a factor of 3 too small. An experiment with the same apparatus but at $R=1.1\times10^{10}$ found $A =  0.68$ K for small $\beta$ \cite{ABN06}, to be compared with the prediction $A \approx 0.33$ K which is  about a factor of 2 too small.  Considering the various approximations made in the derivation of the model, we regard the agreement within factors of 2 or 3 as satisfactory. The predictive  power of the model can be improved by using these fits to obtain empirical values $\delta_{\beta} = 0.00035$ K/s at $R=2.8\times10^8$ and $\delta_{\beta} = 0.0072$ K/s at $R=1.1\times10^{10}$. These values can be used as input for other calculations that depend on these parameters.

\begin{figure}                                                
\centerline{\includegraphics[width=3.25in]{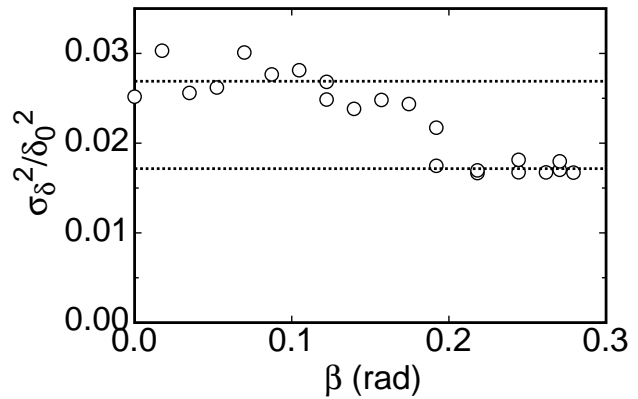}}
\caption{The amplitude variance $\sigma_{\delta}^2$ of $p(\delta)$ normalized by the zero-tilt mean-square amplitude $\delta_0^2$ as a function of tilt angle $\beta$ at $R=2.8\times10^8$. The horizontal dotted lines correspond to the mean values at small and large $\beta$. }
\label{fig:sigma_delta}                                       
\end{figure}

Experimental results for the variance $\sigma_{\delta}^2$ of $\delta$, normalized by the zero-tilt mean-square amplitude $\delta_0^2$,  are shown in Fig.~\ref{fig:sigma_delta} as a function of  $\beta$ for $R=2.8\times10^8$.  Since $\sigma_{\delta}^2/\delta_0^2 \ll 1$, the assumption that the variable damping in Eq.~\ref{eq:lang_tilt_theta} has little effect on average values is justified.  

Diffusion of $\delta$ in a parabolic potential such as that given by Eq.~\ref{eq:potential_lin} with $\beta=0$ was discussed in Ref.~\cite{BA08a}. The probability distribution $p(\delta)$ is a Gaussian function with a variance  
\be 
\sigma_{\delta}^2 = D_{\delta}\tau_{\delta}\ .
\label{eq:sigma}
\ee
In the linearized approximation  given by Eq.~\ref{eq:lang_delta_lin}, the potential is parabolic for small $\epsilon$. The tilt only adds a linear term, so that the sum is a shifted parabola with the same curvature.  Thus, the variance should continue to be given by Eq.~\ref{eq:sigma} even for $\beta > 0$. The parameters $\tau_{\delta}$ and $D_{\delta}$ on the r.h.s of that equation are system parameters that should not depend on the tilt angle. Thus we expect the variance of $\delta$ to be independent of $\beta$. Consistent with that prediction, the experimental values are seen to be nearly independent of $\beta$ for $\beta \stackrel{<}{_\sim} 0.19$. However, they are seen to step down to a smaller constant for larger $\beta$.
The sharp drop in $\sigma_{\delta}^2$ at $\beta\approx 0.19$ rad is inconsistent with a continuous single-well potential.  An analysis of thermistor time-series indicates a sudden change in the behavior of the two thermistors farthest from the plane of the LSC, at approximate orientations $\theta_0\pm \pi/2$, for tilt angles above $\beta \approx 0.19$ rad.  For $\beta = 0$, the fluctuations at $\theta 
\approx \theta_0\pm \pi/2$ are equal to those at $\theta_0$ and only about 15\% less than the maximum which occurs at intermediate angles (see Ref.~\cite{BA07b} for data at $\beta = 0$).  A similar result is found for larger $\beta < 0.19$ rad.  For $\beta > 0.19$ rad, the fluctuations at $\theta \approx \theta_0\pm \pi/2$ become much smaller than the other thermistors by about 2 orders of magnitude. This suggests some sort of change in the LSC structure where the LSC becomes more steady and where perhaps plumes are better aligned with the LSC.

\subsection{Cessation suppression}

\begin{figure}                                                
\centerline{\includegraphics[width=3.25in]{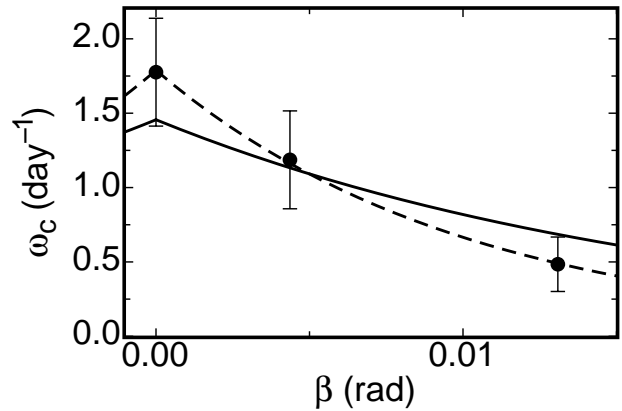}}
\caption{The rate of cessations $\omega_c$ as a function of tilt angle at $R=1.1\times10^{10}$.  Solid line: A fit of $\omega_c \propto \exp(-2\delta_0\delta_{\beta}\beta/D_{\delta})$ to the data with $D_{\delta}/(2\delta_0\delta_{\beta})$ fixed at the value 0.017 based on measurements at $\beta=0$.  Dashed line: a similar fit, with $D_{\delta}/(2\delta_0\delta_{\beta}) = 0.010$ least-squared adjusted.}
\label{fig:cess_rate}                                       
\end{figure}

For small tilt angles, a suppression of cessations in a tilted sample was found in Ref.~\cite{BA06a}, but the rarity of events allowed the collection of only a few data points.  The rate of cessations is shown in Fig.~\ref{fig:cess_rate} for a few small tilt angles at $R=1.1\times10^{10}$, where cessations are more frequent than at $R=2.8\times10^8$.   The error bars represent the expected standard deviation of the mean for Poissonian statistics.  In 12 days of measurements with tilt angles in the range $ 0.017 \mbox{ rad }  \le \beta \le 0.21$ rad, no cessations were found.  

Using the model of diffusion in the potential well given by Eq.~\ref{eq:potential},  a cessation occurs  when $\delta$ diffuses from the bottom of the well up to the top of the barrier at $\delta=0$.  The potential barrier to cessations is then $\Delta V \equiv V(0)-V_{\epsilon}(\bar\epsilon)$, and our system is equivalent to the Arrhenius-Kramers problem. For large potential barriers, the rate of cessations is given by $\omega_c \propto \exp(-2\Delta V/D_{\delta})$ where the proportionality constant depends on the curvatures of the potential near the peak and minimum \cite{Kr40}.  From Eq.~\ref{eq:potential_lin}, the potential barrier in the case of small $\beta$ is 

\be
\Delta V(\beta) = \Delta V(0) + \delta_0\delta_{\beta}\sin\beta + \tau_{\delta}\delta_{\beta}^2\sin^2\beta\ .
\label{eq:delta_V}
\ee

\noindent The curvature of the potential minimum does not vary with $\beta$, but the curvature near the peak does \cite{FN}.   Using the large-barrier-limit prediction of Kramers \cite{Kr40} and the empirical potential of Ref.~\cite{BA08a} (in which a new constant $B=12$ is introduced), the frequency of cessations is given by 

\be
\omega_c = \sqrt{\frac{D_{\delta}\tau_{\delta}}{2\pi\delta_0^2}}\left(B+\frac{2\delta_{\beta}\delta_0\sin\beta}{D_{\delta}}\right)\exp\left(-\frac{2\Delta V(\beta)}{D_{\delta}}\right) \ .
\label{eq:cess_rate_tilt}
\ee

\noindent     The $\sin^2\beta$ term of $\Delta V$ is smaller than the $\sin\beta$ term for $\sin\beta <  0.75$, and the $\sin\beta$ term in the coefficient is smaller than the constant term for $\sin\beta < 0.20$, so for $\beta \ll 0.20$ rad Eq.~\ref{eq:cess_rate_tilt}  simplifies to an exponential decay with $\beta$.  

For $\beta = 0$ the predicted rate of cessations agreed with the experimental result \cite{BA08a} within a factor of order one. However, in order to better test the $\beta$-dependence of the model,
the proportionality 
$\omega_c \propto \exp(-\beta/\beta_0)$ with $\beta_0 = D_\delta/(2 \delta_0 \delta_\beta)$
was fit to the data, adjusting the proportionality constant while fixing $\beta_0 = 0.017$. This value of $\beta_0$ is obtained using the prediction of $\delta_{\beta}$ from Eq.~\ref{eq:dot_delta_beta} and the experimental values for $D_\delta$ and $\delta_0$ that were obtained in Ref.~\cite{BA08a} for  $R=1.1\times10^{10}$. The result is the solid line in the figure. The approximate agreement with the data indicates that the exponential decay rate is consistent (within a factor of two or so)  with the experimental results. Adjusting the proportionality constant as well as $\beta_0$  gave $\beta_0 = 0.010$ and the dashed line in the figure. The value is not very far from the value 0.017 expected on the basis of the independent determinations of the parameters. Although the poor statistics  of the frequency of cessations does not permit a very stringent test of the model, it is noteworthy that both experiment and model indicate that even a very small tilt will almost completely suppress the cessations.  

\subsection{Azimuthal fluctuations}

The stationary probability distribution $p(\dot\theta_*)$ can be found from the steady-state Fokker-Planck equation corresponding to Eq.~\ref{eq:lang_theta_lin} which balances the advection and diffusion of probability.  In Eq.~\ref{eq:lang_theta_lin}, in addition to the explicitly $\beta$-dependent tilt term,  the damping term depends on $\beta$ because $\bar \delta$ is $\beta$-dependent.  The tilt term will be assumed to be small compared to the damping term so it can be ignored.  This is valid in the limit $\sigma_{\dot\theta} \gg \omega_{\beta}^2\beta\sigma_{\theta}$ which is satisfied by the data.  The resulting Fokker-Planck equation gives

\be
-\frac{\dot\theta_0\bar\delta}{\tau_{\dot\theta}\delta_0}p(\dot\theta_*) = \frac{D_{\dot\theta}}{2}\frac{dp(\dot\theta_*)}{d\dot\theta_*}\ .
\ee


\noindent for the statistically stationary state. The solution is $p(\dot\theta_*) \propto \exp(-\dot\theta_*^2/(2\sigma_{\dot\theta}^2))$ with variance

\be
\sigma_{\dot\theta}^2 = \frac{D_{\dot\theta}\tau_{\dot\theta}\delta_0}{2\bar\delta} \ .
\label{eq:sigma_dtheta}
\ee

\begin{figure}                                                
\centerline{\includegraphics[width=3.25in]{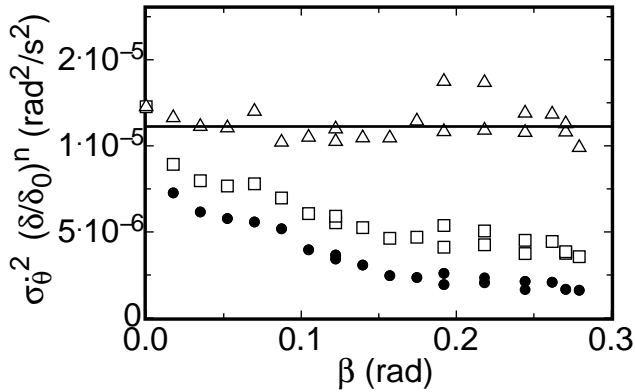}}
\caption{The variance of rotation rate $\sigma_{\dot\theta}^2\times(\bar\delta/\delta_0)^n$ as a function of tilt angles $\beta$ at $R=2.8\times10^8$.  The variance is multiplied by $(\bar\delta/\delta_0)^n$ with several different values of $n$.  Solid circles:   $n=0$, i.e. the bare variance $\sigma_{\dot\theta}^2$.  Open squares:  $n=1$, which is predicted by the model in the linear approximation to yield a constant independent of $\beta$.   Open triangles:  $n=2.29$,  where $n$ is obtained from a best fit of a horizontal line to the data.  }
\label{fig:sigma_dtheta}                                       
\end{figure}

\noindent  The peak of $p(\dot\theta_*)$ was found to be Gaussian and  for $\beta=0$ the predicted variance $\sigma_{\dot\theta}^2$ was found to agree remarkably well with experiment \cite{BA08a}.  The value $\sigma_{\dot\theta}^2$ is plotted as solid circles in Fig.~\ref{fig:sigma_dtheta} for varous tilt angles at $R=2.8\times10^8$. It is seen to decrease with $\beta$.    The $\beta$-dependence of the predicted value is expected to be contained in $\bar\delta$, so the value $\sigma_{\dot\theta}^2\bar\delta/\delta_0$ is expected to be independent of $\beta$. However, the open squares in Fig.~\ref{fig:sigma_dtheta} show that this value also  decreases with $\beta$, albeit more slowly.   Multiplying $\sigma_{\dot\theta}^2$ by $(\bar\delta/\delta_0)^n$, a best fit to a constant gives $n = 2.3$, but we have no explanation for the large error in the predicted exponent.  While this represents a significant disagreement of the model with the data, this width does not affect the overall  phenomenology of the model. 

The power-law relationship between $\sigma_{\dot\theta}^2 $ and $\delta$ is more general and not restricted to tilted samples.   The Gaussian probability distribution and variance from Eq.~\ref{eq:sigma_dtheta} apply instantaneously (with $\bar\delta$ replaced by $\delta$) if the fluctuations in $\dot\theta_0$ are strongly damped so that a stationary $p(\dot\theta_0)$ is reached in a time short compared to that over which  $\delta$ changes.  This is true if $\tau_{\dot\theta} \ll \tau_{\delta}$, which is satisfied for the experimental data.   The near-Gaussian $p(\dot\theta_0)$ implies $|\dot\theta_0| = \sqrt{2/\pi}\sigma_{\dot\theta}$; so it is expected that $|\dot\theta_0| \propto \delta^{-1/2}$.  The measured average relationship was found to be $|\dot\theta_0| \propto \delta^{-1.16\pm 0.06}$ from time series at $\beta=0$  for a wide range of  $R$ \cite{BA06a}.  This gives the exponent for the variance $2.32 \pm 0.12$ which is consistent with the scaling relationship in the tilted sample.  The twist displacement-angle was found to have a similar power-law relationship   $|\theta'(\delta)| \propto \delta^{-1.24}$, which suggests a displacement $\theta'$ may be damped by the same mechanism as $\dot\theta_0$.

The same scaling for $\sigma_{\dot\theta}$ with $\bar\delta$ as the instantaneous relationship suggests that the reduction in $\sigma_{\dot\theta}$ with $\beta$ is indeed due to the increase in $\bar\delta$ as predicted.  In other words the suppression of azimuthal fluctuations is due to the increase in inertial damping with $\delta$ and is unrelated to the orientation-locking term of Eq.~\ref{eq:lang_tilt_theta}.  The reduction in $\sigma_{\dot\theta}$ due to the increased amplitude forcing must be the cause for the suppression of reorientations for small tilt angles that was found in Refs.~\cite{ABN06, BA06a}.  The azimuthal forcing was measured and found not to be strong enough to suppress relatively fast reorientations for small tilt angles despite being able to lock the LSC orientation \cite{BA06a}.  For a large enough tilt, the orientation-locking of the azimuthal equation may play a larger role in suppressing reorientations.

\subsection{Oscillations in a tilted sample}

The linearized azimuthal equation Eq.~\ref{eq:lang_theta_lin} is that of a stochastically forced damped harmonic oscillator.   The power-spectral density $P$  of its solutions is given by

\be
P = \frac{D_{\dot\theta}}{\left[\bar\delta\omega/(\delta_0\tau_{\dot\theta})\right]^2 + \left(\omega_{\beta}^2\sin\beta - \omega^2\right)^2}\ .
\label{eq:power_model}
\ee

\noindent It has a maximum at the resonant frequency $\omega_r = \sqrt{\omega_{\beta}^2\sin\beta - [\bar\delta/(\delta_0\tau_{\dot\theta})]^2/2} $ provided that $\omega_r$ is real. This will be the case when  $\sin\beta > \left[\bar\delta/(\delta_0\tau_{\dot\theta}\omega_{\beta})\right]^2/2$.   This underdamped regime was not attainable in the parameter range of the experiment because the smallest measured value of the damping parameter  was $1/\tau_{\dot\theta} \approx 0.070$ s$^{-1}$ at $R=2.8\times10^{8}$. This is still large compared to the restoring frequency $\omega_{\beta} =  0.011$ s$^{-1}$ even for the maximum possible tilt $\beta=\pi/2$. 

Looking back at the full azimuthal equation, Eq.~\ref{eq:lang_tilt_theta}, one sees that the variable damping term (containing $\delta(t)$ rather than $\bar \delta$)  can add significantly to the excitation of oscillations when the damping occasionally fluctuates to smaller values so that the restoring term temporarily overcomes the damping.  This feature can be incorporated in Eq.~\ref{eq:lang_theta_lin} by adding  the variable damping in the form of  a stochastic term: 

\be
\ddot\theta_* = -\frac{\dot\theta_*}{\tau_{\dot\theta}}[1+\eta(t)] - \omega_{\beta}^2 \sin\beta ~\theta_*+f_{\dot\theta}(t)  \ .
\label{eq:lang_theta_vardamp}
\ee

\noindent Near the stable fixed point one sees from the linearized amplitude equation Eq.~\ref{eq:lang_delta_lin} that the multiplicative noise is $\eta(t) = \epsilon(t)/\delta_0$.   This means $\eta(t)$ is Gaussian distributed with a variance $\sigma_{\eta}^2 = \sigma_{\delta}^2/\delta_0^2$.  It is  not white noise but has the same correlation time $\tau_{\eta} = 2\tau_{\delta}$ as $\epsilon(t)$ \cite{BA08a}.  In other words, this is colored noise of the Ornstein-Uhlenbeck form with correlation function

\be
\langle\eta(t)\eta(t+\tau)\rangle = \sigma_{\eta}^2\exp\left(-\frac{|\tau|}{\tau_{\eta}}\right) \ .
\label{eq:eta_correlator}
\ee

\noindent The power spectrum corresponding to Eq.~\ref{eq:lang_theta_vardamp} can be described by a complicated expression that will not be reproduced here \cite{Gi05}.  A more physically clear result is obtained in the case where $\eta(t)$ represents Gaussian white noise.  In this case the term in the denominator of Eq.~\ref{eq:power_model} due to the damping is reduced by a factor of $1-\sigma_{\delta}^2\tau_{\delta}/(\delta_0^2\tau_{\dot\theta})$ so that the resonance condition is loosened and peaks become stronger with the fluctuation strength $\sigma_{\delta}$.  The value of $\sigma_{\delta}^2\tau_{\delta}/(\delta_0^2\tau_{\dot\theta}) \approx 0.5$  depending on $R$.  This is in a range where factor of two difference in the fluctuation strength allows the system to go from overdamped to effectively undamped, and given the approximate nature of our calculation we cannot with any confidence predict values of the power spectrum or if there will be a resonance peak.  This conclusion applies when we use the full calculation of the power spectrum for colored noise as well.  We can only rely on the prediction that in-phase oscillations are likely to found only at large tilt angles with a resonant frequency that approaches the natural frequency $\omega_{\beta}\sqrt{\sin\beta}$ from below as the resonance peak becomes sharper.

We searched for these azimuthal oscillations experimentally at various tilt angles up to $\beta = 0.26$ rad at $R=2.8\times10^8$.  A complicating factor is that an unrelated  torsional oscillation exists for turbulent convection in cylindrical containers even for $\beta = 0$, in which the orientations of the upper and lower portions of the LSC oscillate out-of-phase around the mid-plane orientation \cite{FA04, FBA08}.  The oscillation predicted due to tilt is expected to have a $z$-independent phase (except possibly for some random fluctuations), that is the orientations of the lower, middle, and upper portions of the LSC are all aligned while they oscillate around the orientation $\theta_{\beta}$.  Since the torsional oscillation is anti-symmetric in $z$, the signal of $\theta_*=\theta_0-\theta_{\beta}$ at the mid-height should only have oscillatory components if the new in-phase oscillatory mode exists.  The out-of-phase oscillation is obtained from differences between the orientations at different heights, such as the signal $\theta' \equiv \theta_0(L/4) - \theta_0(0)$.  

\begin{figure}                                                
\centerline{\includegraphics[width=3.25in]{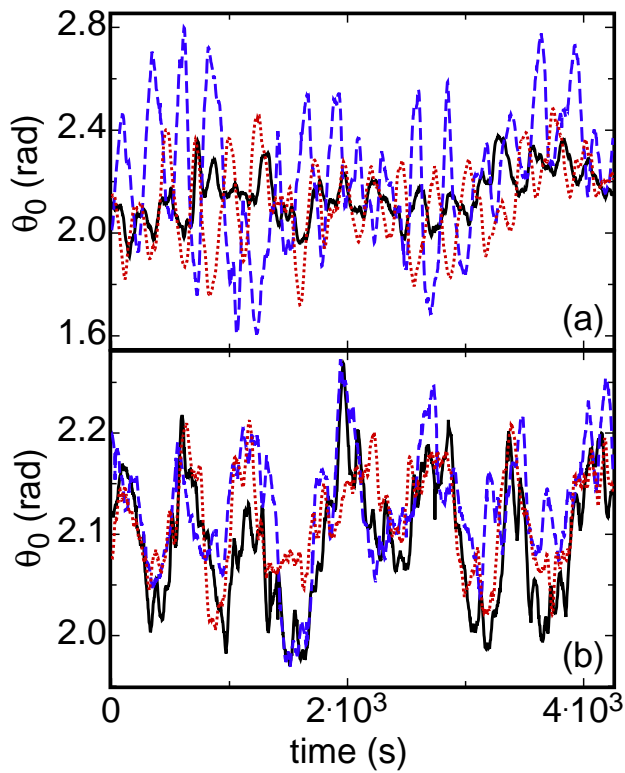}}
\caption{Time series of the orientations $\theta_0$ at the 3 heights at $R=2.8\times10^8$.   (a): for $\beta = 0.09$ rad.  (b):  for $\beta=0.26$ rad.  Dashed line:  $\theta_0(L/4)$.   Solid line:  $\theta_0(0)$.  Dotted line:  $\theta_0(-L/4)$.}
\label{fig:osc_series}                                       
\end{figure}

Time series of the orientations $\theta_0(z)$ at the 3 heights are shown in Fig.~\ref{fig:osc_series}a for $\beta = 0.09$ rad and  Fig.~\ref{fig:osc_series}b for $\beta=0.26$ rad, both at $R=2.8\times10^8$.  For $\beta = 0.09$, the top- and bottom-row orientations oscillate out-of-phase around the middle-row orientation, so this is the torsional oscillation that was observed previously.  For $\beta=0.26$, all 3 rows oscillate in-phase, and at a much lower frequency than the torsional oscillation in Fig.~\ref{fig:osc_series}a.

\begin{figure}                                                
\centerline{\includegraphics[width=3.25in]{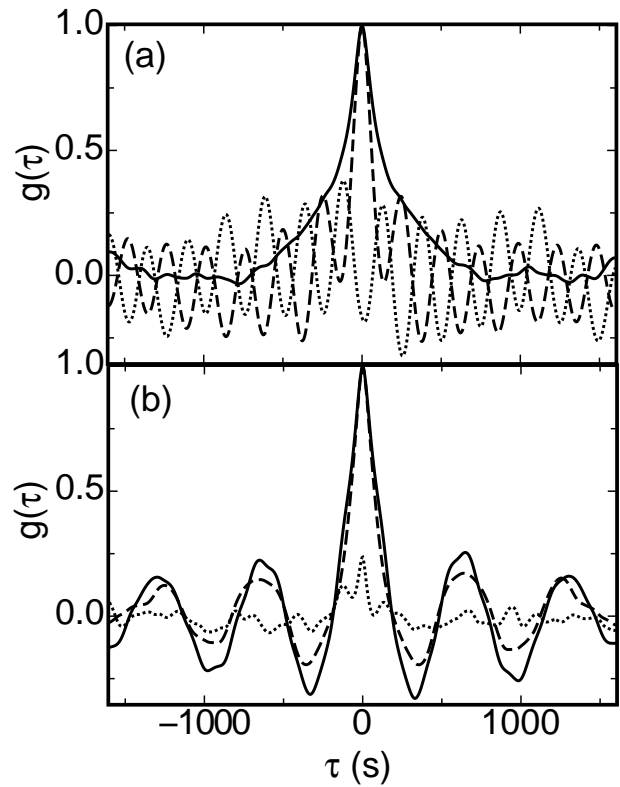}}
\caption{Correlation functions of various orientation differences at $R=2.8\times10^{10}$.   Solid lines:  autocorrelation $g_{*}(\tau)$ of $\theta_*$.      Dashed lines:  auto-correlation $g'(\tau)$ of $\theta'$.  The signal corresponding to the lower-plane orientation relative to the mid-plane orientation  is similar to $\theta'$ so is not shown for clarity.   Dotted  lines:  cross-correlation $g_{tb}(\tau)$ corresponding to the torsional oscillation.  (a): $\beta = 0.09$ rad. (b): $\beta=0.26$ rad.  }
\label{fig:tilt_corr}                                       
\end{figure}

\begin{figure}                                                
\centerline{\includegraphics[width=3.25in]{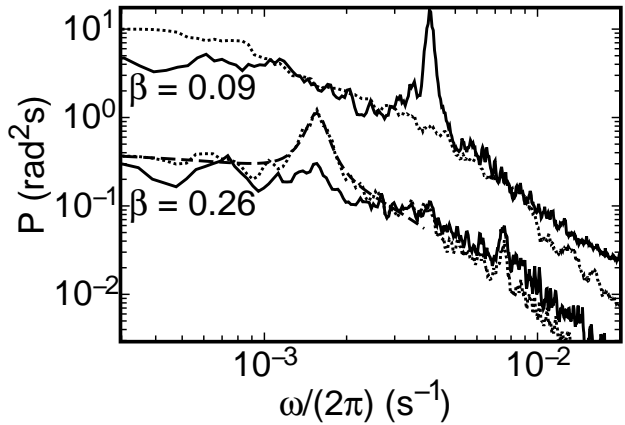}}
\caption{Power spectral densities.  Solid lines:  the Fourier transform of $g_{tb}(\tau)$  corresponding to  the torsional oscillation.  Dotted lines:  The power spectrum of $\theta_*$ corresponding to the in-phase oscillation.  Upper data:  $\beta = 0.09$ rad.  Lower data:  $\beta = 0.26$ rad.  Dashed line:  fit of Eq~\ref{eq:power_spec} to the spectrum for the planar oscillation at $\beta=0.26$ rad.   }
\label{fig:power_spec}                                       
\end{figure}

Autocorrelations $g_{*}(\tau)$ of $\theta_*$ and $g'(\tau)$ of $\theta'$ for $\beta = 0.09$ rad and $\beta=0.26$ rad at $R=2.8\times10^{10}$ are shown in Fig.~\ref{fig:tilt_corr}a and b, respectively.  Also shown is a cross-correlation $g_{tb}(\tau) \equiv \langle[\theta_t(t)-\theta_m(t)][\theta_b(t+\tau)-\theta_m(t+\tau)]\rangle/g_{\theta'}(0)$, which should only oscillate in cases where the LSC undergoes torsional oscillations.  For $\beta=0.09$ rad, both $g_{\theta'}(\tau)$ and $g_{tb}(\tau)$ oscillate as expected for torsional oscillations.   The minimum of $g_{tb}(\tau)$ at $\tau=0$  indicates that the top and bottom rows are out-of-phase, thus showing that this mode corresponds to a torsional oscillation.  The lack of peaks in $g_{\theta_*}$ indicates that there is no discernible in-phase oscillation at $\beta=0.09$ rad.  This plot is qualitatively the same as the results of Ref. \cite{FBA08} for an untilted sample.  For $\beta=0.26$ rad,  the cross-correlation $g_{tb}(\tau)$  is small everywhere, indicating that the torsional oscillation has nearly disappeared. Since the physical origin of the torsional oscillation is unknown, it is difficult to say why this mode is suppressed by tilting the sample.  The new long-period oscillatory mode is apparent in both $g'(\tau)$ and $g_{*}(\tau)$.  

Power spectra of $\theta_*$ for $\beta = 0.09$ rad and $\beta=0.26$ rad at $R=2.8\times10^{10}$ are shown in Fig.~\ref{fig:power_spec}.   Also shown are the power spectra corresponding to the torsional oscillation given by the Fourier transform of $g_{tb}(\tau)$.  The peak of each power spectrum was fit by a Lorentzian function 

\be
P(\omega) = \frac{P_0}{1 + (\omega-\omega_0)^2/\sigma_p^2}  
\label{eq:power_spec}
\ee

\noindent  with an empirically chosen exponential background for the in-phase oscillation and power-law background for the torsional oscillation.    

\begin{figure}                                                
\centerline{\includegraphics[width=3.25in]{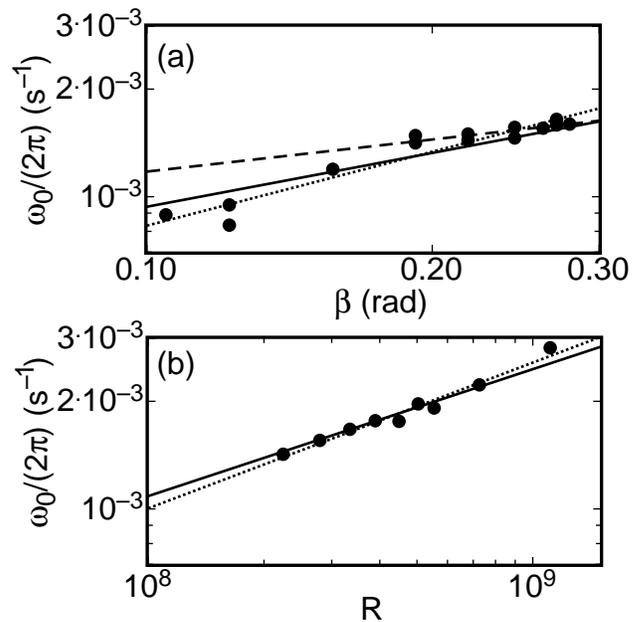}}
\caption{The resonant frequency $\omega_0$ obtained from the peak of the power spectrum of $\theta_*$ on log-log scales.  (a): Data as a function of tilt angle $\beta$ at $R=2.8\times10^8$.  Solid line:  A fit of $\omega_0 = \omega_{\beta}\sqrt{\sin\beta}$ to the data. Dotted line:  a best power law fit allowing the exponent to vary gives an exponent of 0.69.  Dashed line:  a best power law fit to data for $\beta > 0.19$ rad allowing the exponent to vary gives an exponent of 0.30.  (b):  Data as a function of $R$ at $\beta = 0.26$ rad.  Solid line:  A power-law fit with the exponent fixed at the predicted value 0.355.  Dotted line: a best power law fit allowing the exponent to vary gives an exponent of 0.41.}
\label{fig:osc_freq}                                       
\end{figure}

Since the analytic form of the power spectrum and any expression for its peak frequency are complicated in the case of the in-phase oscillations, we compare $\omega_0$ to the model prediction for the natural frequency $\omega_{\beta}\sqrt{\sin\beta}$ even though the resonant frequency tends to be shifted to a somewhat lower value than the natural frequency.   The agreement between the resonant frequency and natural frequency is expected to be better when the resonant peak is sharper, which is found to occur for larger $\beta$ and smaller $R$.  A plot of the fit values $\omega_0$ as a function of tilt angle is shown in Fig.~\ref{fig:osc_freq}a for $R=2.8\times10^8$.  A fit by the equation $\omega_0  = \omega_{\beta}\sqrt{\sin\beta}$ is shown as the solid line. The expected $\sqrt{\sin{\beta}}$ dependence is reasonably consistent with the data. The coefficient was found to be  $\omega_{\beta} = 1.9\times10^{-2}$ s$^{-1}$, within a factor of two of the predicted value $\omega_{\beta} = 1.1\times10^{-2}$ s$^{-1}$.  Since the resonance peaks are small at small $\beta$, they are likely in the range where the resonant frequency is shifted below the natural frequency, so it might be appropriate to only fit f the data for large $\beta > 0.19$ rad which gives an exponent of 0.30.  On the other hand, a best power law fit to all of the data gives an exponent of 0.69, so the data are consistent with a wide range of power laws between those two values.
    
A plot of $\omega_0$ for various $R$ at $\beta=0.26$ rad is shown in Fig.~\ref{fig:osc_freq}b. Over the $R$-range of the data the Nusselt number can be described well by ${\cal N} = 0.159R^{0.290}$ \cite{NBFA05}.  Thus from Eq.~\ref{eq:omega_beta} one finds $\omega_0 \propto R^{0.355 }$.  A power-law fit with this exponent gave the solid line in the figure, which is in excellent agreement at smaller $R$ where the resonance peaks are sharpest.  The fit gave a prefactor of $9.8\times 10^{-6}$ s$^{-1}$ which is again within a factor of two of the predicted value $5.7\times10^{-6}$ s$^{-1}$.  A best power law fit to the data gives an exponent of 0.41, quite close to the model prediction.  Since the measured frequency of in-phase oscillations is within a factor of two of the predicted value with scalings in $\beta$ and $R$ that are consistent with the model, it is clear that the observed oscillations are due to the restoring force created by tilting the sample as predicted.

\begin{figure}                                                
\centerline{\includegraphics[width=3.25in]{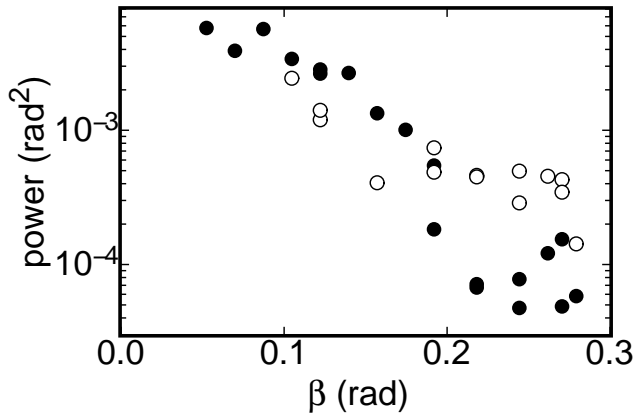}}
\caption{The power $\pi P_0 \sigma_p$ equal to the integral of the peak of the power spectrum  corresponding to the azimuthal oscillations.   Open circles:  planar oscillation.   Solid circles:  torsional oscillation.   }
\label{fig:osc_power}                                       
\end{figure}

The mean-square amplitude of the oscillations is equal to the power given by the integral over the peak of the power spectrum, i.e. to $\int_{peak}P(\omega)d\omega = \pi P_0 \sigma_p$.  This power is plotted in Fig.~\ref{fig:osc_power} for both the torsional and the in-phase oscillations.  The power of the torsional oscillation drops sharply at $\beta\approx 0.19$. At larger $\beta$  the planar oscillation has the larger amplitude.   This transition was clearly visible in time series of $\theta_0$, where only the torsional oscillation was visible for $\beta \stackrel {<}{_\sim} 0.19$, while only the in-phase oscillation was visible for larger $\beta$.  This transition occurs at the same $\beta$ as the drop in $\sigma_{\delta}^2$ (see Fig.~\ref{fig:sigma_delta}), further suggesting that there may be a structural change of the LSC at this tilt angle.

\section{Pressure forcing in cylinders with non-circular cross section}
\label{sect:noncircular}

Containers with strong azimuthal asymmetry are known to align the LSC plane parallel to the longest diameter \cite{DE01, ZSX07}, but other effects on the dynamics have not been well-studied.  
Here we model the LSC in a container with a non-circular horizontal cross section, corresponding to an azimuthally varying diameter $D(\theta)$. This model predicts a restoring force that is a consequence of pressure gradients due to the non-circular cross section of the sample which vanish in the circular limit.  This restoring force leads to a preferred orientation of the LSC aligned with the longest diameter(s) of the container.  If the azimuthal asymmetry is large enough the model predicts the existence of oscillations that are in phase over the entire height of the sample.

The methods of Ref.~\cite{BA08a} are used to produce estimates for volume-averaged terms that can be added as a perturbation to the Langevin equation Eq.~\ref{eq:lang_theta}.  As usual, the model assumes a singe-roll LSC. It is assumed that the azimuthal average of $D(\theta)$ is equal to $L$ and that $D(\theta)$ remains close to $L$ at all $\theta$ so that the LSC has an aspect ratio approximately equal to one, independent of $\theta_0$.  We will only consider the flow in a central band near the LSC orientation where $\theta \approx \theta_0$. This simplifies the problem, and it is expected to be the region where the side-wall-generated pressure gradient has the largest influence. The velocity in the plane of the LSC in this region is assumed to be

\be
u_{\phi} = \frac{\omega_{\phi} r_{\phi} L}{D(\theta)}
\label{eq:vel_profile}
\ee

\noindent where $r_{\phi}$ is the radius in the plane of the LSC and $\omega_{\phi}$ is the angular turnover frequency.  For the circular cross section the radial profile is supported by the measurements of Refs.~\cite{QT01a, QT01b}.  The factor $L/D$ makes the velocity at the edge of the bulk the same at all $\theta$, which is the case if the viscous boundary-layer width and profile are independent of $\theta_0$.  This also makes $\omega_{\phi}$ uniform.   The LSC is forced into this nearly circular loop due to the pressure from the side walls, which requires the pressure term from the NS equation to provide the centripetal acceleration

\be
\dot u_{r} = \frac{u_{\phi}^2}{r_{\phi}} = -\frac{d \mathcal{P}}{dr_{\phi}}
\ee

\noindent where $\mathcal{P}$ is the pressure per unit density.  Using Eq.~\ref{eq:vel_profile} this can be integrated to get the pressure field 

\be
\mathcal{P}= \frac{\omega_{\phi}^2 L^2 r_{\phi}^2}{2D(\theta)^2} \ .
\ee

\noindent Now the $\theta$-derivative is taken to get the azimuthal acceleration: 

\be
\dot u_{\theta} = -\frac{1}{r}\frac{d\mathcal{P}}{d\theta} = \frac{\omega_{\phi}^2 L^2 r_{\phi}^2 D'(\theta)}{D(\theta)^3 r} 
\ee

\noindent  where $D'(\theta) \equiv dD/d\theta$ can have a significant fractional variation with $\theta$ despite $D(\theta)$ remaining close to $L$.  The volume-averaged azimuthal acceleration $\ddot\theta_g$ which can be added to the Langevin equation Eq.~\ref{eq:lang_theta} can be obtained by canceling out $r_{\phi}/r$ to get

\be
\ddot \theta_{g} = \frac{\dot u_{\theta}(\theta_0)}{r} = \frac{\omega_{\phi}^2 L^2 D'(\theta_0)}{D(\theta_0)^3} \ . 
\label{eq:geo_forcing}
\ee

When added as a perturbation to Eq.~\ref{eq:lang_theta}, the geometry effect can be expressed in terms of diffusion in a potential given by

\be
V_g \equiv -\int \ddot\theta_g d\theta = \frac{\omega_{\phi}^2L^2}{2D(\theta_0)^2}\ .
\label{eq:potential_geo}
\ee

\noindent  There are two fixed points where $D'(\theta_0)=0$. The longest diameter corresponds to the lowest potential, and thus its orientation is the  preferred one for the LSC circulation plane. Qualitatively this is similar to the experimental observations for samples with a square or rectangular cross section  \cite{DE01, ZSX07} where the alignment is along the longest diagonal. Since $D(\theta)$ must be $\pi$-periodic, there is at least one longest diameter in the cell and an even number of opposite preferred orientations along these diameters.  Containers with rectangular cross-sections have four preferred orientations.  It is interesting to note that transitions between opposing preferred orientations would correspond to the {\it reversals}  that have often been discussed in the literature \cite{NSSD01,FGL05,Be05} but that are not prevalent in the circular cross-section limit where azimuthal diffusion in the absence of potential extrema leads to equal probability for all LSC orientations.

The forcing came out of the pressure term as a result of the assumed azimuthal variation of the internal flow field.  If on the other hand, $u_{\phi}$ is independent of $\theta$ and $\omega_{\phi}$ varies with $\theta$, then there is no azimuthal force.  However, for such a profile to satisfy no-slip boundary conditions in a non-cylindrical container would require the boundary-layer profile to vary with $\theta$.  This variation would have to be drastic if the variation in $D(\theta)$ is comparable to or larger than the boundary-layer width.

An alternate derivation  of the azimuthal forcing that does not depend on the internal velocity profile is obtained using a simple physical model of a vector force-balance at the wall.  The pressure at the side wall provides an acceleration that can be broken down into two orthogonal components: the centripetal acceleration $\dot u_{r} = 2u_{\phi}^2/D$, and the azimuthal acceleration $\dot u_{\theta}$.  Since the pressure is normal to the wall surface, simple geometry gives the ratio of these acceleration comonents to be $D(\theta_0)/D'(\theta_0)$, so 
\be
\ddot\theta_0 =  \frac{2\dot u_{\theta}}{D} =\frac{4u_{\phi}^2 D'(\theta_0)}{D^3(\theta_0)} \ .
\ee
In this description the result of a forcing proportional to $D'(\theta)$ is not dependent on an assumption about the internal flow field or the assumption of small deviation from azimuthal invariance.  However it still depends on  $u_{\phi}$, and is equal to the previous result using Eq.~\ref{eq:vel_profile} to get $u_{\phi} = \omega_{\phi}L/2$ at the wall.  In this description it is clear that the pressure of the side wall -- whose gradient  is no longer directed towards the sample center for a non-circular cross-section --  provides the acceleration to align the flow with the longest diameter.

The probability distribution of the orientation can be obtained from the steady-state Fokker-Planck equation in the strong-damping limit and is

\be
p(\theta_0)\propto \exp\left(-\frac{V_g}{\tau_{\dot\theta}D_{\dot\theta}}\right)
\label{eq:fokker_planck_pressure}
\ee

\noindent The rate of switchings between different preferred orientations for various cross sections can be found by directly applying Kramers' prediction.

\subsection{Oscillations in containers with an elliptic cross section}	

Let us take for example a sidewall shape that is slightly elliptical so the diameter varies as $D(\theta)/L = 1 + \varepsilon\cos(2\theta)$ where $\varepsilon \ll 1$ so the aspect ratio remains close to 1.   Then $D'(\theta)/L = -2\varepsilon\sin(2\theta_0)$ where $\theta_0 = 0$ and $\pi$ are the preferred orientations and $2\varepsilon$ is the difference between the minor and major axes.  Taking the small-angle approximation $\sin(2\theta_0) \approx 2\theta_0$ and adding this forcing due to the elliptical sidewall to the linear equation Eq.~\ref{eq:lang_theta_lin} gives the stochastically driven linear harmonic-oscillator equation

\be
\ddot\theta_0 = -\frac{\dot\theta_0}{\tau_{\dot\theta}} - 4\varepsilon\omega_{\phi}^2\theta_0 +f_{\dot\theta}(t) \ .
\label{eq:slho}
\ee

\noindent Here the ellipticity parameter $\varepsilon$ plays a similar role to the tilt angle $\beta$ in Eq.~\ref{eq:lang_theta_lin}. The qualitative difference between the two cases is that the container geometry is not expected to affect the LSC strength because it does not enter into the $\delta$ equation.   The resonant frequency of the deterministic system is 

\be
\omega_r^g = \sqrt{4\varepsilon\omega_{\phi}^2 - \frac{1}{2\tau_{\dot\theta}^2}}
\ee

\noindent so the system becomes underdamped above a critical ellipticity $\varepsilon_c = 1/(8\omega_{\phi}^2\tau_{\dot\theta}^2)$. For $R=1.1\times10^{10}$, for instance, we find $\varepsilon_c \simeq 0.16$. But as for the case for the tilted sample, the inclusion of the factor $\delta/\delta_0$ in the damping term of Eq~\ref{eq:slho} will lead to oscillations for $\epsilon < \epsilon_c$. A detailed consideration would lead to a result similar to Eq.~\ref{eq:lang_theta_vardamp}.

The linear oscillator equation applies when the orientation is confined to a small region due to the asymmetry. If the asymmetry is weak, so that the potential barrier $\Delta V_g$ equal to the difference between the maximum and minimum of the potential is not large compared to the effective fluctuation energy $D_{\dot\theta}\tau_{\dot\theta}/2$, then the orientation can switch between potential minima.   If the fluctuation energy is much larger, switchings become frequent and the effect of the potential is minimal.  For the elliptical container, the well depth is $\Delta V_g = 2\varepsilon\omega_{\phi}^2/(1-\varepsilon^2)^2$, so to have bounded oscillations, the ellipticity must be greater than $\varepsilon_r \approx \tau_{\dot\theta} D_{\dot\theta}/(2\omega_{\phi}^2) = 2\tau_{\dot\theta}^3 D_{\dot\theta}\varepsilon_c$. For $R=10^{10}$ for instance this yields $\varepsilon_r \approx  0.02\varepsilon_c \approx 0.003$. Using the measured values from Ref. \cite{BA08a} one finds a very weak dependence on $R$.  This small but non-zero ellipticity required to lock the orientation is consistent with experiments in non-circular containers which found a locked orientation, as well with measurements for nominally circular containers -- which must have some small effective ellipticity -- that found that the orientation meanders. 
When $\varepsilon \gg \varepsilon_r$, then the rate of reversals between the two orientations along the longest diameter is given by the Kramers result 
\be
\omega_r = \frac{2\sqrt{2}\varepsilon}{\pi\tau_{\dot\theta}}\exp\left(-\frac{\varepsilon}{\varepsilon_r}\right) \ .
\label{eq:cess_rate_elliptical}
\ee

\subsection{Containers with rectangular cross section}

Common experimental geometries besides circular cylinders have been samples with square or rectangular cross-sections.  Although the model should work best for smooth cross-sections with small $D'(\theta)/L$, it is instructive to apply it to the rectangular case.   With a square cross-section, for example, the shortest and longest diameters can be compared to get an effective ellipticity $\varepsilon_{eff} \approx (\sqrt{2}-1)/(\sqrt{2}+1) \approx 0.29$. This is larger than the predicted  $\varepsilon_c$ required to get oscillations.  Because of the approximate nature of the prediction, at best we can say that there is a good chance of finding azimuthal oscillations in a square cell due to the pressure of the side wall, and that reversals should be extremely rare or absent.

\begin{figure}                                                
\centerline{\includegraphics[width=3in]{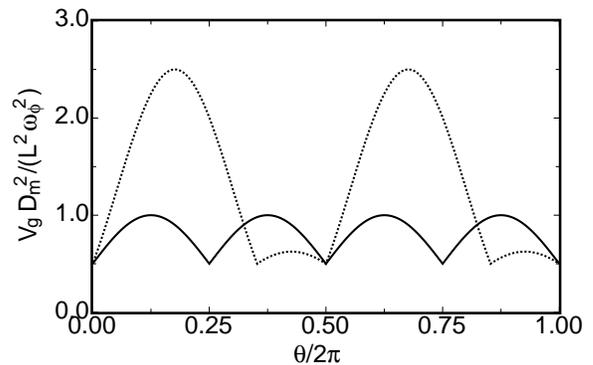}}
\caption{The azimuthal potentials $V_g$ for containers with rectangular cross-sections.  Solid line: $\Lambda  =1$, i.~e.~a square cross-section.   Dotted line:  $\Lambda=2$. }
\label{fig:rect_potential}                                       
\end{figure}

The Fokker-Planck analysis on rectangular cross-section containers leads to an unusual result for $p(\theta_0)$.  Consider a rectangular container of side lengths $a_+$ and $a_-$ so that the longest diameter is $D_m \equiv (a_+^2 + a_-^2)^{1/2}$ with cross-section aspect ratio $\Lambda \equiv a_+/a_-$. Let $\theta =0$ along a diameter and $D_{\pm}$ correspond to a wall with length $a_{\pm}$.  Then the diameter can be expressed as 
\be
D_\pm(\theta) = \frac{D_m}{|\cos\theta| + \Lambda^{\pm1}|\sin\theta|}
\label{eq:D_of_theta}
\ee
provided $|\theta_{\pm}| < \pi -2\arctan \left(\Lambda^{\mp 1}\right)$.
This result applies to all four corners but is mirror-imaged at alternate corners.  Examples of the full potential $V_g$ given by Eqs.~\ref{eq:potential_geo} and \ref{eq:D_of_theta} are shown in Fig.~\ref{fig:rect_potential}.

The variation in diameter is at least as large as that of the square cross-section, so reversals will usually be strongly suppressed in a rectangular container. However, switchings between nearby corners may occur for sufficiently large $\Lambda$. For $\Lambda$ close to one, a small-angle expansion of $p(\theta_0)$ around the longest diameters is a good approximation.  A first-order expansion of the diameter around a corner gives 
\be
D_{\pm}(\theta_0) \approx (1-\Lambda^{\pm 1}|\theta|)D_m\ .
\ee
The steady-state Fokker-Planck equation Eq.~\ref{eq:fokker_planck_pressure} gives 
\be
p_{\pm}(\theta_0) \propto \exp\left(-\frac{2|\theta_0| \Lambda^{\pm 1}\omega_\phi^2 L^2}{D_m^2 \tau_{\dot\theta}D_{\dot\theta}}\right) \ .
\ee
This distribution has a width of $2\varepsilon_r D_m^2/(L^2 \Lambda^{\pm 1})$ which is on the order of 1 degree. One sees that a sample with a square cross section ($\Lambda = 1$) gives an exponential peak near a corner with equal decay rates on the two sides adjacent to the corner.
A rectangular cross-section ($\Lambda > 1$) also gives an exponential peak, but with a faster decay rate on the longer side.  In comparison, an elliptical cross section gives a Gaussian peak due to the parabolic shape around the longest diameter.  Thus a test of this model would be to measure $p(\theta_0)$ in containers with rectangular  cross-sections.  A polygon with  corners having more obtuse angles would be expected to give wider distributions of $p(\theta)$ that might be easier to resolve, and increase the rate of switchings between nearby corners.   

To determine the rate of switching between adjacent corners, we compare the potential barrier $\Delta V_s$ between the corners with the diffusive energy $\tau_{\dot\theta}D_{\dot\theta}/2$.  The potential barrier between adjacent corners is 
\be
\Delta V_s = \frac{\omega_{\phi}^2 L^2}{2D_m^2 \Lambda^{\pm2}}
\ee
where $\Lambda^2$ is used for the barrier between the closer corners and $\Lambda^{-2}$ is used for the barrier between the further corners. The potential barrier is equal to the diffusive energy  when $\Lambda\approx L/(2D_m\sqrt{\varepsilon_r}) \approx 6.3$  for $R=10^{10}$ and assuming $L = D_m/\sqrt{2}$. Thus for larger $\Lambda$ the rate of switchings between adjacent corners becomes frequent, while for smaller $\Lambda$ the rate of switchings drops off rapidly with the usual exponential form proportional to $\exp[-2\Delta V_s/(\tau_{\dot\theta}D_{\dot\theta})]$.  For large $\Lambda$, the orientation may not be locked into a single corner and the rate of switching between nearby corners could become appreciable.  However, such large $\Lambda$ values are far from a nearly uniform-diameter cross-section; thus it is unclear how reliable our approximations are in this limit.

\subsection{Comparison to prediction for an ellipsoidal container}

The deterministic model of the LSC presented in Ref.~\cite{RPTDGFL06} used the Boussinesq equations with slip boundary conditions for ellipsoidal samples.  Convection was due to an applied uniform vertical temperature gradient along one axis of the ellipsoid. It is apparent that this geometry is not very closely related to any that have been studied experimentally so far.  The main virtue of the  model is that it allowed an exact solution for the velocity field, but dissipation had to be introduced {\it a posteriori} and phenomenologically.   For certain parameter ranges the model leads to an in-phase azimuthal oscillation for $\varepsilon > \varepsilon_c \simeq 0.02$ of the LSC around a preferred orientation in which its circulation plane is aligned on average along the longest horizontal diameter of the container.
The value $\varepsilon_c = 0.02$ is  below our stochastic-model prediction for underdamped oscillations.
For $\varepsilon > 0.04$, the deterministic model also found various chaotic modes.
A common feature of our stochastic model and the deterministic one of Ref.~\cite{RPTDGFL06} is that in both cases the oscillations are a consequence of pressure gradients due to the non-circular cross section of the sample. However, in our model they are those of a damped oscillator driven by a stochastic force.

The experimental results of Ref.~\cite{RPTDGFL06} included an azimuthal oscillation mode with a frequency that matched the turnover frequency of the LSC.  The measurement was made only near the top plate. Thus it could not be determined whether there was a phase difference between the top and bottom of the flow.  This was reported to be consistent with the in-phase oscillation predicted by the deterministic model presented in the same paper, provided  that  $\varepsilon$ was used as a tunable parameter.  It seems more likely that the observed oscillations were the torsional oscillation mode reported in Ref.~\cite{FA04}, which is already known to have the same frequency without the need for a tunable parameter.  Physically, there was no reason to assume the large ellipticity required to obtain oscillations synchronous with the turnover frequency because the container had a nominally circular cross section.  The torsional oscillation is known to occur in azimuthally symmetric containers, and because of the symmetry, it does not oscillate around a fixed orientation but rather around the mid-plane orientation which meanders \cite{FBA08}. The experiments of Refs.~\cite{RPTDGFL06} and \cite{ZSX07} found that the azimuthal oscillations occurred around a fixed orientation in the experiment, but that may well be due to locking of the LSC orientation  due to various small asymmetries which occurs to some extent in all samples \cite{BA06b}.

\section{Other perturbations}

\subsection{Asymmetric heating}

It was found in Ref.~\cite{KH81} that asymmetric heating can align the LSC orientation.   A small asymmetric heating effect was measured in Ref.~\cite{BA06b} and found to produce alignment of the LSC in the direction of a small horizontal temperature gradient at the top plate. Asymmetric heating enhances the buoyancy of the boundary layers near the hotter side of the sample, and the dynamic effects on the LSC are expected to be similar to the effect of tilting the sample. Thus one expects the potential well of $\delta$ to become deeper, leading to a suppression of cessations.  

Consider a temperature difference $\delta T$ over a length $L$ horizontally across a plate such that the hot side is in the direction $\theta_c$ so the LSC is forced in this direction.  We volume-average forces using the same methods as Ref.~\cite{BA07a} to obtain the effects on both $\theta_0$ and $\delta$.  The forcing in the direction of $\theta_c$ in terms of acceleration is \cite{BA06b} 
\be
\langle\dot u'\rangle_V = \frac{g\alpha \delta T}{3\pi \mathcal{N}} \ .
\ee
One vector component gives the azimuthal forcing 
\begin{eqnarray}
\ddot\theta_c &=& \frac{\langle\dot u'\rangle_V}{\langle r\rangle} \sin(\theta_c-\theta_0)\\
&=& \frac{g\alpha \delta T }{\pi \mathcal{N}L} \sin(\theta_c-\theta_0)
\end{eqnarray}
which acts to align the flow and can be added as a perturbative term to Eq.~\ref{eq:lang_theta}.  The other component acts to strengthen the LSC when it is aligned with the asymmetry and gives 

\begin{eqnarray}
\dot\delta' &=& \frac{\delta \langle\dot u'\rangle_V}{U}  \cos(\theta_c-\theta_0) \\
&=&\frac{g\alpha \delta T L \delta}{3\pi\mathcal{N} Re \nu} \cos(\theta_c-\theta_0) 
\end{eqnarray}

\noindent which can be added as a perturbative term to Eq.~\ref{eq:lang_delta}.

\subsection{The Coriolis force}
\label{sect:coriolis}

The experiment of Ref.~\cite{BA06b} studied the effects of Earth's Coriolis force on the LSC.  The model of the Coriolis force presented in Ref.~\cite{BA06b} was equivalent to the strong-damping limit of Eq.~\ref{eq:lang_theta} with additional forcing terms that both align the LSC and cause a net azimuthal rotation of the LSC plane. A more complete analysis of the integration of the two models was presented in Ref.~\cite{BA08a}.

A significant qualitative difference between the several asymmetries that we have discussed is that the Coriolis force and non-circular side walls only add terms to the azimuthal equation Eq.~\ref{eq:lang_theta} while the heating and tilt asymmetries add terms to both the azimuthal equation Eq.~\ref{eq:lang_theta} and the amplitude equation Eq.~\ref{eq:lang_delta}.  In general the Coriolis force always acts perpendicular to a flow direction. Thus it can deflect a flow but not strengthen or weaken it.  Since the Coriolis force only affects the azimuthal motion and not the LSC strength, it should not affect cessations which we found to be suppressed by the increase of the amplitude in the case of a tilted sample. This was studied in Ref.~\cite{BA06b} by tilting the container slightly and observing that the peaks or minima of measured parameters did not occur at $\beta=0$, but at a slightly shifted angle due to the interplay between Coriolis and tilt forces in one set of experiments, and the interplay between a heating asymmetry, Coriolis, and tilt forces in another experiment.  The conclusion was that the measured azimuthal forcing as a function of $\theta_0$ and the parameters corresponding to $p(\theta_0)$ were found to be affected by all three mechanisms.  However, $\delta$, $R_e$, $\sigma_{\dot\theta}$, and the frequency of reorientations were only affected by the heating and tilt asymmetries.  This is in  agreement with the stochastic model, since $\delta$, $R_e$, $\sigma_{\dot\theta}$, and the frequency of reorientations should only be affected by an asymetry that also enhances the LSC strength.  The data of Ref.~\cite{BA06b} were taken for only very weak asymmetries (the apparatus was stationary on the surface of the Earth, so the Coriolis force measured was due to Earth's rotation), so those data support the conclusion that a weak Coriolis force does not suppress reorientations.  This conclusion has not been tested in systems with stronger asymmetries.  A large enough asymmetry could change the structure of the LSC, in which case the models presented here assuming a single-roll LSC would not apply.

\section{Summary and Conclusions}

In a previous paper \cite{BA08a} we developed a model for the large-scale circulation in turbulent Rayleigh-B\'enard convection in cylindrical containers. It consisted of two stochastic ordinary differential equations, one for a measure $\delta$ of the strength (Eq.~\ref{eq:lang_delta}) and the other for the azimuthal orientation $\theta_0$ (Eq.~\ref{eq:lang_theta}) of the flow. This model preserved the rotational invariance of the ideal cylindrical system. In many, if not most, real physical systems, including idealized laboratory experiments, this invariance is broken by a variety of perturbations. In the present paper we considered the effects of tilting the cylinder axis relative to gravity, deviation from a circular cylinder cross section, asymmetric heating of the top plate, and the influence of the Coriolis force due to rotation. We did this by adding physically motivated terms to either or both of the two equations as required by the nature of the perturbation. Predictions based on this perturbed model were compared with experiments for tilted samples, and generally there was agreement within factors of order unity between the measured and predicted quantities. Predictions for the influence of the Coriolis force due to Earth's rotation had been compared with experiment in a previous paper \cite{BA06b}.

A particularly interesting result of this work is that asymmetries can be categorized based on how they affect the flow dynamics.  Those that influence the buoyancy of the thermal boundary layers, such as a tilt or asymmetric heating, increase the strength of the LSC. They lead to an additive term of the $\delta$-equation. We show that they suppress cessations and azimuthal fluctuations because they deepen the potential well corresponding to the $\delta$-equation.  
Other asymmetries can result in the addition of a perturbative term only to the $\theta_0$ equation. Examples include the influence of a Coriolis force due to rotation of the sample, and the influence of deviations from a circular cross  section of the sample. These asymmetries align the LSC, but do not affect its strength. As a consequence, for instance,  they do not change the frequency of  cessations of the LSC.

Another interesting feature of the model is that it predicts the generation of an azimuthal oscillation of the LSC circulation plane when the tilt is sufficiently strong. These oscillations are due to an azimuthal restoring force that arises when the sample LSC orientation deviates from the tilt direction. This case was investigated in detail by experiment, and the predicted oscillations, with frequencies close to the predicted values, were indeed observed. It is important to note that this oscillation is expected to be in phase over the entire sample length, in contrast to a quite different torsional mode of unknown origin that is known to exist even for the untilted sample \cite{FBA08}. Whereas the torsional mode has a period that is synchronous with the LSC turnover time, this new mode generally has a longer period that depends on the tilt angle and is not related to the turnover time.  

An oscillatory mode, coherent over the entire sample height, is predicted as well for a sufficiently large eccentricity of an elliptical cylinder cross section. In this case the oscillations are predicted to be due to a  pressure gradient induced by the deviation from circular symmetry. Such an asymmetry is predicted to lead to two preferred orientations, with the LSC circulation plane aligned with the longer axis of the ellipse. Oscillations can occur about one or the other of these alignments. When the eccentricity is small, the LSC orientation is expected to diffuse in the neighborhood of one or the other of the two orientations, perhaps with occasional reversals of the flow direction.

A long-term goal of the approach that we have taken would be to apply these results to natural systems, such as flow in Earth's outer core or the solar convection zone which are spherically symmetric shells with a Coriolis force.  These flow structures are much more complicated than a single LSC roll, but while the asymmetries align these circulations, we have shown that when these asymmetries are small they do not affect the LSC strength and should not affect the frequency of cessations. However, further experimentation should be carried out  to establish whether stronger asymmetries that align the LSC can also affect its stability to cessations.

\begin{figure}
\centerline{\includegraphics[width=3in]{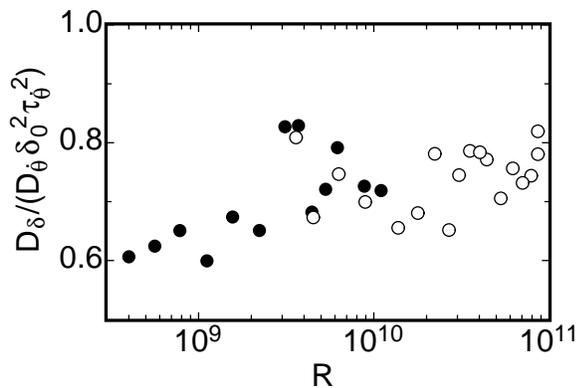}}
\caption{The dimensionless ratio between diffusivities $D_{\delta}/(D_{\dot\theta}\delta_0^2\tau_{\dot\theta}^2)$ for $\beta=0$.  Solid symbols: data from the $L=24.76$ cm sample.  Open symbols: data from a similar sample with $L=50.61$ cm, referred to as the large sample in Ref.~\cite{BNA05}.  The ratio is expected to be 1 if fluctuations have the same strength in both the $\phi$ and $\theta$ coordinates.  }
\label{fig:diff_ratio}                                    
\end{figure}

\section{Appendix:  Comparison of diffusivities}

The basic model Eqs.~\ref{eq:lang_delta} and \ref{eq:lang_theta} involves the two stochastic forces $f_{\delta}(t)$ and $f_{\dot\theta}(t)$ that lead to diffusivities  $D_{\delta}$ and $D_{\dot\theta}$ of $\delta$ and $\dot \theta$. Here we compare these diffusivities.  One might hypothesize that the thermal fluctuations in the system are isotropic, so the fluctuations would have the same strength both in the direction of the LSC velocity and in the azimuthal motion.  From Eq.~\ref{eq:temp_profile} the variance of temperature fluctuations is given by 
\begin{eqnarray}
\sigma_T^2&=&\langle (d\delta)^2\cos^2(\theta-\theta_0) + (d\theta_0)^2 \delta^2\sin^2(\theta-\theta_0) \rangle \\ 
&\approx& [(d\delta)^2 + (d\theta_0)^2\bar\delta^2 ]/2\ .
\end{eqnarray}
If the $\delta$ and $\theta_0$ terms contribute equally to the total fluctuations, then equipartition gives $(d\delta)^2 = \bar\delta^2(d\theta_0)^2$.  Using the diffusivity relationships found in Ref.~\cite{BA08a} 
and simplifying to the $\beta=0$ case where $\bar\delta = \delta_0$ leads to

\be
\frac{D_{\delta}}{D_{\dot\theta} \delta_0^2 \tau_{\dot\theta}^2}  = 1 \ .
\label{eq:diff_ratio}
\ee

\noindent We plot the l.~h.~s.~of this equation in Fig.~\ref{fig:diff_ratio} using the experimental values from Ref.~\cite{BA08a} for $\beta=0$. The ratio is slightly less than 1 with little variation over our experimental range of $R$, so by this measure the thermal fluctuations in the direction of the LSC are smaller than in the azimuthal direction, if only slightly.  For the purposes of the model, the fluctuations can be considered nearly isotropic, so use of Eq.~\ref{eq:diff_ratio} reduces the number of free parameters of the model by 1.

\section{Acknowledgments}

This work was supported by Grant DMR07-02111 of the US National Science Foundation.  We thank Chris Henley for his insightful analysis of our data.



\begin{thebibliography}{10}
\bibitem{FN}In Ref.~\cite{BA08a} the potential Eq.~\ref{eq:potential} was found to be inconsistent  with the data for $p(\delta)$ at small $\delta \alt 0.5\delta_0$. Thus an empirical potential was introduced  for which, consistent with the data,  $dV/d\delta = -BD_{\delta}\delta_0$ for small $\delta$ with $B = 12$ for $R=1.1\times10^{10}$.
\end{thebibliography}

\end{document}